# Metallicity and Anomalous Hall Effect in Epitaxially-Strained, Atomically-thin RuO$_2$ Films

Seung Gyo Jeong[1,†,*], Seungjun Lee[2,†,*], Bonnie Lin[3], Zhifei Yang[1,2], In Hyeok Choi[5], Jin Young Oh[6], Sehwan Song[7], Seung wook Lee[5], Sreejith Nair[1], Rashmi Choudhary[1], Juhi Parikh[1], Sungkyun Park[7], Woo Seok Choi[6], Jong Seok Lee[5], James M. LeBeau[3], Tony Low[2,*], and Bharat Jalan[1,*]

[1]Department of Chemical Engineering and Materials Science, University of Minnesota−Twin Cities, Minneapolis, Minnesota 55455, USA
[2]Department of Electrical and Computer Engineering, University of Minnesota, Minneapolis, Minnesota 55455, USA
[3]Department of Materials Science and Engineering, Massachusetts Institute of Technology, Cambridge, Massachusetts 02139
[4]School of Physics and Astronomy, University of Minnesota, Minneapolis, Minnesota 55455, USA
[5]Department of Physics and Photon Science, Gwangju Institute of Science and Technology (GIST), Gwangju 61005, Republic of Korea
[6]Department of Physics, Sungkyunkwan University, Suwon 16419, Republic of Korea
[7]Department of Physics, Pusan National University, Busan 46241, Korea

[†] Equally contributed

[*]Corresponding authors:
Seung Gyo Jeong, Email: jeong397@umn.edu
Seungjun Lee, Email: seunglee@umn.edu
Tony Low, Email: tlow@umn.edu
Bharat Jalan, Email: bjalan@umn.edu

**Author Contributions:** S.G.J., S.L., T.L., and B.J. conceived the idea and designed the experiments. S.G.J., S.N., and J.P. grew the films. S.G.J., Z.Y., S.N., and R.C. characterized the samples and performed electrical transport measurements. S.L., and T.L. performed the first-principles calculations. B.L. performed the electron ptychography and corresponding analysis under the supervision of J.M.L. J.Y.O. performed X-ray absorption spectroscopy and ellipsometry under the supervision of W.S.C. S.G.J., S.L., T.L., and B.J. wrote the manuscript. All authors contributed to the discussion and manuscript preparation. B.J. directed the overall aspects of the project.

**Competing Interest Statement:** The authors declare no competing interests.

**This PDF file includes:**

    Main Text
    Figures 1 to 3

**Abstract**

The anomalous Hall effect (AHE), a hallmark of time-reversal symmetry breaking, has been reported in rutile $RuO_2$, a debated metallic altermagnetic candidate. Previously, AHE in $RuO_2$ was observed only in strain-relaxed thick films under extremely high magnetic fields (~50 T). Yet, in ultrathin strained films with distinctive anisotropic electronic structures, there are no reports, likely due to disorder and defects suppressing metallicity thus hindering its detection. Here, we demonstrate that ultrathin, fully-strained 2 nm $TiO_2$/$t$ nm $RuO_2$/$TiO_2$ (110) heterostructures, grown by hybrid molecular beam epitaxy, retain metallicity and exhibit a sizeable AHE at a significantly lower magnetic field (< 9 T). Density functional theory calculations reveal that epitaxial strain stabilizes a non-compensated magnetic ground state and reconfigures magnetic ordering in $RuO_2$ (110) thin films. These findings establish ultrathin $RuO_2$ as a platform for strain-engineered magnetism and underscore the transformative potential of epitaxial design in advancing spintronic technologies.

**Introduction**

Ruthenium dioxide (RuO$_2$) has emerged as a focal point in condensed matter physics due to its intriguing magnetic properties and potential applications in spintronics—a rapidly growing field that leverages electron spin for advanced information processing and storage. The magnetic ground state of RuO$_2$ has however been the subject of intense debate, with studies yielding conflicting results. Early investigations using neutron (2017) and resonant X-ray diffraction techniques (2019) reported small magnetic moments in both bulk and thin films of RuO$_2$ at room-temperature (1, 2). These findings further suggest antiferromagnetic ordering in RuO$_2$ which has long been considered a Pauli paramagnet. However, the most recent studies (2024) employing muon spin rotation (µSR) spectroscopy and neutron scattering observed negligible magnetic moments, thereby challenging earlier claims and casting doubt on the magnetism in RuO$_2$ in both its bulk and thin film forms (3, 4).

Alongside these efforts, growing evidence now points to the emergence of altermagnetism in RuO$_2$ *thin* films—a newly proposed class of magnetic materials with compensated magnetic orders yet exhibiting time-reversal symmetry breaking behaviors (5-13). Unlike conventional antiferromagnets, the magnetic symmetry in an altermagnet is preserved by time-reversal and rotational symmetry operation, giving rise to alternating spin-split bands in momentum space commonly of *d*-, *g*- or *i*-wave characters. Angle-resolved photoemission spectroscopy (ARPES) has confirmed such spin-splitting in the band structure of RuO$_2$ films (5, 14). Magneto-optical measurements have detected responses consistent with altermagnetic behavior (15), and spin-splitting torque experiments have demonstrated spin-dependent phenomena in RuO$_2$ thin films (7-11). Additionally, observations of the thermal Hall effects (12) and anomalous Hall effect (AHE) in thick and/or doped RuO$_2$ (110) films (6, 11, 16) further support its altermagnetic nature. Notably, symmetry-sensitive optical second-harmonic generation (SHG) studies have also uncovered time-reversal symmetry breaking in thin $RuO_2$ epitaxial films below 500 K (13), implicating a magnetic octupole as the relevant order parameter for altermagnetism (17, 18).

These conflicting results have ignited vigorous debate over the actual magnetic ground state of $RuO_2$, while also underscoring discrepancies in the sample quality across different studies. For thin films, variations in thickness, epitaxial strain, and defect concentration are likely contributing factors to these inconsistencies. In particular, epitaxial strain, arising from the lattice mismatch between the $RuO_2$ film and $TiO_2$ substrate, can profoundly influence the material's electronic and magnetic properties by modifying orbital energies and spin interactions, as evidenced by resonant elastic X-ray scattering studies (19). Nonetheless, acquiring comprehensive transport data on fully strained $RuO_2$ films remains challenging, largely owing to the large lattice mismatch with conventional substrates and difficulties in synthesizing high-quality epitaxially-strained $RuO_2$ films while maintaining their metallicity.

In this article, we address these challenges using the hybrid molecular beam epitaxy (hMBE) technique (20, 21) to grow ultrathin, epitaxial, fully-strained $RuO_2$ (110) films. Films were capped with an epitaxially-grown insulating 2 nm $TiO_2$ (110) film to prevent it from surface contamination upon air exposure. Through a combination of high-resolution X-ray diffraction, multislice electron ptychography, X-ray absorption and photoemission spectroscopies, magneto-transport measurements, and density functional theory (DFT) calculations, we demonstrate that $RuO_2$ films remain metallic down to the unit cell limit and exhibit a pronounced anomalous Hall effect (AHE). Notably, the epitaxial strain stabilizes a non-compensated magnetic ground state, where AHE signatures were observed under a 9 T magnetic field, contrary to previous observations of AHE in thicker $RuO_2$ films requiring extremely high magnetic fields (~50 T) (6, 11). Our findings furnish decisive evidence for the pivotal role of epitaxial strain in both inducing and tuning magnetic order in ultrathin RuO$_2$ films, thus resolving recent debates over its magnetic ground state in thin film form.

**Results and Discussion**

To build a robust foundation for subsequent electronic transport analysis, we first provide a thorough structural characterization of our RuO$_2$ films, highlighting their excellent structural quality and low disorder, thus facilitating metallicity down to a thickness of 2 unit cell. Figure 1a presents a schematic of our sample structure, comprising ultrathin 2 nm TiO$_2$/$t$ nm RuO$_2$/2 nm TiO$_2$ on a TiO$_2$ (110) substrate. The thin TiO$_2$ buffer layer was incorporated to ensure a clean and well-defined bottom interface, minimizing substrate-induced variations and ensuring consistent growth across different samples. Representative reflection high-energy electron diffraction (RHEED) patterns (Fig. 1b) recorded after the growth of a 2 nm TiO$_2$/1.9 nm RuO$_2$/2 nm TiO$_2$ on a TiO$_2$ (110) sample show streaky RHEED patterns accompanied by the Kikuchi lines along the [1$\bar{1}$0] (upper panel) and [001] (lower panel) crystal directions, indicating the high crystalline quality. The clear RHEED oscillations observed during growth (Fig. S1) confirm precise thickness control of the TiO$_2$ layer and reveal the layer-by-layer growth mode. Atomic force microscopy (AFM) images of the heterostructure (Fig. 1c) reveal an atomically smooth surface with a root mean square roughness ($S_q$) of 175 pm, further corroborating the high-quality growth. The presence of well-defined step-terrace structures is consistent with epitaxial growth and streaky RHEED patterns.

High-resolution X-ray reflectivity (XRR) and diffraction (XRD) θ-2θ scans (Figs. 1d and 1e) demonstrate that the RuO$_2$ films adopt a (110)-oriented out-of-plane lattice structure. The observed thickness oscillations in both XRR and XRD confirm atomically sharp surfaces and interfaces within the heterostructures. The periodicity of the oscillations matches the total thickness of the layers. Notably, the Kiessig's fringes in XRR persist up to a high angle of ~8° (Fig. 1d) which enables precise determination of each layer thickness including RuO$_2$ down to 0.4 nm by fitting the experimental data (solid lines). At room temperature, tetragonal lattice constants of bulk RuO$_2$ are $a = b = 4.492$ Å and $c = 3.106$ Å, while those of bulk TiO$_2$ are $a = b = 4.594$ Å and $c = 2.959$ Å. For the (110) plane, the lattice mismatches between RuO$_2$ and TiO$_2$ are defined as $-4.7\%$ along [001] and $+2.3\%$ along [1$\bar{1}$0] (22). XRD reciprocal space mapping (RSM) results (Fig. S2) reveal that the RuO$_2$ films remain fully-strained for thicknesses up to 7.9 nm, with anisotropic strain relaxation observed at 9.1 nm. Importantly, the TiO$_2$ buffer and capping layer effectively suppress surface relaxation in RuO$_2$, extending the critical thickness for strain relaxation compared to RuO$_2$ grown directly on the substrate (22). Together, these structural characterizations affirm the high-quality epitaxial growth and the fully-strained film in 2 nm TiO$_2$/$t$ nm RuO$_2$/2 nm TiO$_2$ on a TiO$_2$ (110) heterostructures for $t \leq 7.9$ nm.

To complement these global structural characterizations, we conducted a cross-sectional multislice electron ptychography image (23) along the [001] zone axis of a rutile heterostructure with $t = 2.1$ nm, as shown in Fig. 1f. Because the reconstructed phase of the object correlates with the atomic number, Ru, Ti, and oxygen ions are identified as shown in the overlay on the right side of Figure 1. The RuO$_2$/TiO$_2$ interfaces, marked by horizontal dotted lines in Fig. 1f, are atomically sharp at both the upper and lower boundaries. Based on these boundaries, the RuO$_2$ layer thickness (2.0 nm) determined from the ptychographic reconstruction aligns closely with XRR analysis (2.1 nm). This observation aligns with the persistent Kiessig's fringes observed in the XRR data shown in Fig. 1d, further confirming the atomically-smooth interfaces in the heterostructure. Additionally, X-ray photoemission and X-ray absorption spectroscopy (Figs. S3 and S4) confirm the robust +4 oxidation states of Ti and Ru atoms within the RuO$_2$/TiO$_2$ (110) heterostructures. Together, these structural and chemical analyses highlight the high crystallinity, atomically sharp interfaces, and smooth surfaces of these $RuO_2$/$TiO_2$ heterostructures.

We measured the electrical resistivity ($\rho$) of 2 nm $TiO_2$/$t$ nm $RuO_2$/2 nm $TiO_2$ heterostructures as a function of $t$ to further evaluate structural disorder in metallic $RuO_2$. Resistivity measurements at 300 K, performed using the Van der Pauw (VdP) method, revealed metallic behavior in the $t$ = 0.8 nm film, with $\rho$ = 332 µΩcm. However, the $t$ = 0.4 nm heterostructure exhibited 4-point resistance values exceeding the measurement range (> 5 MΩ), indicating insulating behavior. This abrupt transition between 0.4 and 0.8 nm is likely caused by discontinuities in the ultrathin film due to the single-unit-cell (~0.3 nm) step-terrace surface. To benchmark the metallicity of our ultrathin $RuO_2$ sample, we compared the 300 K resistivity values with those of other few-nanometer-thick metallic systems including metallic oxides, elemental metals, and two-dimensional materials reported in previous studies (Fig. 1g). Our $RuO_2$ films demonstrated the lowest $\rho$, particularly at thicknesses below 2 nm, emphasizing the superior structural and electronic properties. The sample with $t$ = 0 nm (i.e., the $TiO_2$ layers only) shows no measurable resistivity, effectively ruling out any contribution from the $TiO_2$ layer to the overall resistivity. This confirms that the observed resistivity in the $RuO_2$ films is solely due to the $RuO_2$ layer and highlights the metallic nature of the ultrathin $RuO_2$ films. We also note that the thickness dependence of resistivity indicates the absence of an interface conductance channel between $RuO_2$ and $TiO_2$, suggesting no electronic or chemical reconstruction at the interface.

We now turn to the discussion of temperature-dependent magneto-transport measurements for representative samples: one fully-strained ($t$ = 1.7 nm) and one partially relaxed ($t$ = 9.1 nm) $RuO_2$ film. Figures 2a and 2b show the temperature-dependent $\rho$ for these films, measured using a Hall bar device with current along the [1$\bar{1}$0] direction (whose schematic is shown in the inset of Fig. 2a). For the $t$ = 1.7 nm film (Fig. 2a), $\rho$ at 300 K aligns with the results from VdP method (Fig. 1g). Hall measurements conducted with both VdP and Hall bar devices (Fig. S5) show consistent results, indicating no significant degradation from the Hall bar fabrication process. The resistivity remains metallic down to 100 K (d$\rho$/d$T$ > 0), with a slight upturn below 100 K, likely due to localization effect (24). In contrast, the $t$ = 9.1 nm film (Fig. 2b) displays the metallic temperature dependence to the lowest temperature, previously observed in thicker strain-relaxed RuO$_2$ films (6, 11, 24). For the $t$ = 1.7 nm film (Fig. 2c), the magnetic ($H$)-field dependent Hall resistivity ($\rho_H$) exhibits non-linear behavior at $T \leq 15$ K whereas the $t$ = 9.1 nm film (Fig. 2d) shows linear behavior at all temperature accompanied by a sign change in the Hall slope below 300 K. Similar nonlinear Hall behavior was observed for current direction along [001] (Fig. S6). It should be noted that the relaxed 9.8 nm RuO$_2$/TiO$_2$ (110) film exhibits a nonlinear Hall effect only at ~ 50 T (11), so its absence in our 9.1 nm film is not surprising given the limited field range (± 9 T) applied in this study. Consequently, we focus on the non-linear Hall features in the fully-strained $t$ = 1.7 nm film. Nonlinear Hall effects can arise from multiple carriers' conduction or from the anomalous Hall effect (AHE). Based on the failure of applying multiple conduction model to our experimental results and DFT-based calculations on strained RuO$_2$ thin films, we attribute the observed non-linearity to the AHE (discussed in the SI, Note S1, Fig. S7). We also performed an oxygen annealing study to exclude the possibility of oxygen vacancies contributing to conduction and nonlinearity (discussed in the SI, Note S1, Fig. S8).

To quantitatively analyze the AHE in ultrathin RuO$_2$, we extracted the AHE contribution from $\rho_H(H)$ using the equation $\rho_H(H) = R_oH + \rho_{AHE} \tanh(cH)$, where $R_o$, $\rho_{AHE}$, and $c$ are the ordinary Hall coefficient, AHE resistivity, and a fitting parameter describing the $H$-field dependence of AHE, respectively. The ordinary Hall effect was removed by linear fitting of the high-field data near 9 T. The left panel of Fig. 2e shows the fitting results (solid line) for the measured $\rho_H(H)$ (circles) of the $t = 1.7$ nm film at 1.8 K. The fit reveals a saturated AHE with $\rho_{AHE} \approx 0.3$ μΩ cm (right panel of Fig. 2e), comparable with values reported for thicker and doped RuO$_2$ films (6, 11, 16) and ultrathin SrRuO$_3$ superlattices (25). Figure 2f further exhibits the temperature-dependent AHE conductivity ($\sigma_{AHE}$) with $\sigma_{AHE} = -\rho_{AHE}/(\rho_{AHE}^2 + \rho^2)$ for the $t = 1.7$ nm film where $\rho$ is the measured longitudinal resistivity. Experimental measurements and fitting analyses from other fully-strained samples are shown in Fig. S9, exhibiting the same nonlinear behavior. Figure 2g summarizes $\sigma_{AHE}$ at 1.8 K as a function of $t$ ($\leq 4$ nm). Interestingly, as $t$ decreases from 3.9 nm to 0.9 nm, $\sigma_{AHE}$ increases. However, at $t = 0.8$ nm, $\sigma_{AHE}$ sharply decreases, suggesting a potential dimensionality-driven shift in transport and/or magnetic properties, rather than by epitaxial strain alone.

We performed DFT calculations to investigate the nonlinear behavior, which revealed that epitaxial strain can stabilize long-range magnetic order in RuO$_2$ films grown on TiO$_2$ (110), even without the Hubbard $U$ parameter. As shown in Fig. 3a, the averaged local magnetic moment of Ru atoms, $|\mu_{Ru}^A - \mu_{Ru}^B|/2$, increases significantly with (110) in-plane epitaxial strain ($\varepsilon$) at $U = 0$. Here A and B refer to the two Ru-sublattices in the RuO$_2$ unit cell. For fully strained RuO$_2$ ($\varepsilon = 1$), $|\mu_{Ru}^A - \mu_{Ru}^B|/2$ reaches 0.366 μ$_B$, with a net magnetic moment of 0.156 μ$_B$/cell. This non-compensated magnetic order emerges from symmetry breaking introduced by the (110) epitaxial strain, specifically the disruption of [$C_2 \| C_4 t$] symmetry (13), and persists even without spin-orbit coupling, distinct from relativistic effects (26). In this context, [$C_2 \|$ $C_4 t$] symmetry describes the two-fold spin rotation in conjunction with a combined operation of four-fold crystal rotational and half-translation. Figure 3b shows the relative total energies of nonmagnetic (NM), ferromagnetic (FM), and non-compensated antiferromagnetic (AFM) states as a function of $\varepsilon$ without $U$. At $\varepsilon = 1$ (fully strained state), non-compensated AFM represents the magnetic ground state, while FM becomes energetically favorable at $\varepsilon = 0.9$, thus demonstrating strain tuning as an effective experimental knob in controlling magnetic order in RuO$_2$. It should further be noted that at $\varepsilon = 1$, non-compensated AFM becomes even more stable with the addition of a small value of $U \geq 0.4$ (Fig. S10).

Spontaneous magnetization in an itinerant (anti)ferromagnet can often be described by the Stoner criterion, DOS($E_F$)$I_0 > 1$, where DOS($E_F$) is the density of states at Fermi energy for the non-magnetic states, and $I_0$ is the Stoner parameter (typically, $I_0 = 0.5$ eV for 4$d$ metals (27)). Figure 3c highlights strain-induced modifications of DOS($E_F$) in RuO$_2$, with narrow peaks near the Fermi level primarily attributed to the $t_{2g}$-$d_{x^2-y^2}$ states of Ru 4$d$ orbitals (22). In-plane epitaxial strain shifts these peaks towards the Fermi level, triggering Fermi surface instability and magnetic ordering in strained RuO$_2$. While this picture offers an intuitive understanding based on DFT, a complete quantitative analysis involving spin susceptibility at the relevant wave vector, is beyond the scope of this work. Nevertheless, our calculations demonstrate that epitaxial strain alone can stabilize a magnetic ground state in RuO$_2$, providing a straightforward route to intrinsic magnetism in strained RuO$_2$ films, distinct from magnetism induced by hole doping or including Hubbard $U$ in bulk RuO$_2$ (36) (See the SI, Note S2, Figs. S10, S11).

To further investigate the AHE in the magnetic ground state, we evaluated the magnetic anisotropy energy (MAE) and AHE of fully strained RuO$_2$ at $U = 0$ (Figs. 3d–3f). Figure 3d shows that the MAE, plotted against the spin orientation angle ($\theta$), reaches a minimum at $\theta = 0°$ (in-plane [001] direction), identifying this direction as the magnetic easy axis. The MAE increases with $\theta$, reaching a value of 1.1 meV/cell at $\theta = 90°$ (out-of-plane [110] direction), which is lower than that of reported bulk RuO$_2$ where a sizable $U$ value is required to stabilize magnetism (27). This reduced MAE, combined with non-compensated magnetic order, likely facilitates the observed AHE at relatively low magnetic fields compared to relaxed 9.1 nm

RuO$_2$ films. The MAE of 1.1 meV/cell corresponds to ~13 K, matching the temperature for the onset of AHE in Fig. 2g but well below the ~500 K magnetic transition temperature observed in SHG studies (13). The MAE represents the energy barrier for reorienting the magnetization direction and is generally smaller than the exchange energy associated with the magnetic transition temperature. Consequently, AHE is expected to emerge at temperatures well below 500 K. Additionally, owing to magnetic fluctuations, the AHE only appears below temperatures set by the MAE (~13 K).

We further examined how $\sigma_{AHE}$ in strained RuO$_2$ depends on the energy near the Fermi level for various spin orientation angles $\theta$, as shown in Fig. 3e. At $\theta = 0$, $\sigma_{AHE}$ vanishes due to symmetry but appears immediately with small $\theta$, as summarized in Fig. 3f. The negative $\sigma_{AHE}$ aligns with experimental observations for 0.8 nm < $t$ < 4 nm films (Fig. 2g), with the relatively modest experimental values suggesting a small $\theta$ at 9 T of $H$-field. Notably, $\sigma_{AHE}$ increases with $\theta$ and saturates at $\theta = 90°$, potentially explaining the $\sigma_{AHE}$ enhancement observed between $t$ = 0.9 and 3.9 nm in Fig. 2g. In this thickness range, while the in-plane lattice constants of RuO$_2$ remain consistent with TiO$_2$, interface-induced octahedral distortions in ultrathin RuO$_2$ may further modify $\theta$ and/or MAE, thereby amplifying $\sigma_{AHE}$.

In summary, we have investigated the AHE in atomically thin, metallic RuO$_2$ heterostructures, revealing their intricate magnetic and transport properties. Utilizing precision oxide heterostructure synthesis via hMBE, advanced characterization techniques, and DFT calculations, we uncovered a non-compensated magnetic order stabilized by epitaxial strain. Our experiments demonstrated a pronounced AHE in epitaxially-strained RuO$_2$ layers (≤ 4 nm) at low temperatures and under modest $H$-fields (9 T), standing in contrast to the behavior of strain relaxed thicker films. Theoretical analysis linked the AHE to epitaxial strain-driven symmetry breaking and octahedral distortions, underscoring their impact on magnetic anisotropy and conductivity. Collectively, these findings demonstrate that strained ultrathin RuO$_2$ films can serve as a fertile platform for investigating emergent magnetic ground states and offer promising avenues for developing spintronic technologies leveraging metallic and magnetic functionalities.

**Materials and Methods**

**Hybrid MBE and structural characterizations**

Epitaxial $RuO_2/TiO_2$ heterostructures were synthesized using an oxide hybrid molecular beam epitaxy (hMBE) system (Scienta Omicron) on $TiO_2$ (110) single-crystal substrates (Crystec). The heterostructures comprised $RuO_2$ layers sandwiched between buffer and capping $TiO_2$ layers. Prior to growth, substrates were cleaned sequentially with acetone, methanol, and isopropanol, followed by baking at 200 °C for 2 hours in a load lock chamber. Just before deposition, substrates were annealed at 300 °C with oxygen plasma for 20 minutes to achieve a clean surface. Oxygen plasma during all growth processes was operated at 250 W forward power with a chamber pressure of $5 \times 10^{-6}$ Torr.

For $RuO_2$ growth, the metal-organic precursor $Ru(acac)_3$ was thermally evaporated using a low-temperature effusion cell (MBE Komponenten) maintained at 170–180 °C. Titanium tetraisopropoxide (TTIP, 99.999%, Sigma-Aldrich, USA) was used for $TiO_2$ growth, introduced via a gas inlet system equipped with a linear leak valve and a Baratron manometer, with a beam equivalent pressure (BEP) of $3 \times 10^{-7}$ Torr. The substrate temperature was kept at 300 °C for both $RuO_2$ and $TiO_2$ layer depositions. Post-growth, the samples were cooled to 120 °C in an oxygen plasma environment to minimize oxygen vacancy formation. Film surfaces were monitored in situ using RHEED, (Staib Instruments) before, during, and after growth. Crystalline quality, film thickness, roughness, and strain state were analyzed using high-resolution X-ray diffraction (XRD, Rigaku SmartLab XE). Surface morphologies were examined with atomic force microscopy (AFM, Bruker Nanoscope V Multimode 8) using peak-force tapping mode.

**X-ray photoelectron spectroscopy**

To confirm the oxidation state of Ti and Ru elements in epitaxial $RuO_2/TiO_2$ heterostructures with a comparison with pristine TiO$_2$ substrates, we measured core-levels X-ray photoelectron spectroscopy (XPS, Physical Electronics VersaProbe III) at Ti 2$p$, Ru 3$d$, and O 1$s$ edges with a monochromatic Al Kα X-ray source (1486.6 eV). We used a flood gun to prevent photoemission-induced surface charge effects. XPS spectrum was measured by an energy step size of 0.05 eV and a pass energy of 55 eV. The binding energy of all XPS results was calibrated using the binding energy of C–C bonding (C 1$s$, 284.4 eV) from the adventitious carbon on the surface. We note that the C 1$s$ peak is very close to the Ru 3$d_{3/2}$. Nevertheless, we expect that the C 1$s$ peak is much more pronounced than the Ru 3$d_{3/2}$ peak because RuO$_2$ is capped by 2 nm of TiO$_2$ capping layer. More importantly, we could not measure any additional Ti and Ru states deviating from the +4 oxidation state within measurement limits.

**X-ray absorption spectroscopy**

Ru $L$ and O $K$ edges XAS spectra of RuO$_2$/TiO$_2$ thin films are measured at the 6A beamline of the Pohang Light Source using the total electron yield (TEY) method. The measurements were conducted at room temperature and used linear horizontal (LH) polarization, where the electrical field direction of the X-rays is parallel to [1$\bar{1}$0] crystal direction of RuO$_2$/TiO$_2$ (110) heterostructures. We rotated the samples for the incoming X-rays by an incidence angle, 90° (surface normal) and 60° (grazing). To observe possible different oxidation states at the interfaces, we measured both a 2 nm TiO$_2$/17 nm RuO$_2$ film and a superlattice structure (2 nm RuO$_2$ and 1 nm TiO$_2$ repeated five times, total 15 nm thick), as shown in Fig. S4. In both thin films, we did not observe any states deviating from Ru$^{4+}$ and Ti$^{4+}$ within measurement limits, consistent with XPS.

**Electrical transports**

Temperature-dependent resistance, magnetoresistance, and Hall resistance measurements were conducted in both Van der Pauw and Hall bar geometries with aluminum wire bonding in a physical property measurement system (PPMS, Dynacool, Quantum Design, USA). To ensure the current directions, Hall bar devices were fabricated along two distinct in-plane directions, [001] and [110], using photolithography followed by reactive ion etching, Etching was performed in a reactive ion etcher (Advanced Vacuum) with a mix of 30-sccm $O_2$ and 1.5-sccm $SF_6$ with 200 W power in 20 mTorr pressure..

**Multislice electron ptychography**

Cross-sectional samples for electron microscopy were prepared along [001] through conventional wedge polishing, followed by Ar ion milling. A Thermo Fisher Scientific Themis Z aberration-corrected scanning/transmission electron microscope operating at 300 kV was used to collect the 4D-STEM datasets for multislice electron ptychography. The incident electron probe was defined by a convergence semi-angle of approximately 26 mrad, an overfocus of 10 nm, and a beam current of 15 pA. The 4D-STEM dataset was collected with a 128x128 pixel Electron Microscope Pixel Array Detector (EMPAD) (28). with a scan step size of 0.0377 nm/px, a diffraction pixel size of 0.69 mrad/px, and a dwell time of 1 ms. Sample thickness was determined via position-averaged convergent beam electron diffraction (PACBED) (29) to be approximately 20 nm. Ptychographic reconstructions were performed using a modified version of the fold_slice (23) fork of the PtychoShelves (30) software package. The multislice reconstruction engine was a GPU-accelerated maximum-likelihood solver with multiple state mixtures to account for (partial) coherence of the probe (31-33). A total of 16 incoherent probe modes were considered. In total, 760 iterations were used for the reconstruction, broken into four major stages. In each step, the probe remained fixed for the first 25 iterations. The regularization parameter began at 1.0 and gradually decreased to 0.1 by the final stage. The sample thickness was maintained constant at 20 nm throughout all iterations, but with the number of slices varying from 4 at the first stage and increasing to 20 slices for the final stage. Sample surface damage was removed by considering and averaging only the middle five slices.

**Density functional theory calculation**

We performed first-principles calculations based on density functional theory (34) as implemented in the Vienna *ab initio* simulation package (35). The projector augmented wave potentials (36, 37) were used to describe the valence electrons, and the plane-wave kinetic energy cutoff was chosen to be 500 eV. The exchange-correlation function was treated by the generalized gradient approximation (GGA) of Perdew-Burke-Ernzerhof (38). To describe (110) oriented RuO$_2$, we aligned three lattice vectors of RuO$_2$ along [001], [1$\bar{1}$0], and [110] directions, respectively. The theoretical equilibrium lattice constants of bulk RuO$_2$ were calculated to be $a_{[001]}^{RuO_2} = 3.119$ Å and $a_{[1\bar{1}0]}^{RuO_2} = a_{[110]}^{RuO_2} = 6.394$ Å. To consider epitaxial strain effects of RuO$_2$ thin films from TiO$_2$, we employed lattice constants of $a_{[001]}^{TiO_2} = 2.970$ Å and $a_{[1\bar{1}0]}^{TiO_2} = 6.588$ Å, corresponding to theoretical equilibrium lattice constants of bulk TiO$_2$. The in-plane lattice constants of epitaxially-strained RuO$_2$ were determined as $a_{[001]} = (1-\varepsilon)a_{[001]}^{RuO_2} + \varepsilon a_{[001]}^{TiO_2}$ and $a_{[1\bar{1}0]} = (1-\varepsilon)a_{[1\bar{1}0]}^{RuO_2} + \varepsilon a_{[1\bar{1}0]}^{TiO_2}$, where $\varepsilon$ represents degree of the epitaxial strain. Then, we used a fully relaxed out-of-plane lattice constant of $a_{[110]}$ at given $\varepsilon$. For example, at $\varepsilon=1$, $a_{[110]}$ was calculated to be 6.527 Å. The corresponding Brillouin zones were sampled with an 20 × 10 × 10 *k*-grid mesh. To evaluate anomalous Hall conductivity ($\sigma_{AHE}$), we constructed the Wannier Hamiltonian using the Wannier90 package (39, 40), characterized by Ru *d* and O *p* orbital projections. Then, we calculated the intrinsic $\sigma_{AHE}$ by employing the Berry curvature formula. To obtain converged results, we used a fine *k*-grid mesh of 200 × 200 × 200.

**Acknowledgements**

Film synthesis and structural characterization (S.G.J and B.J.) was supported by the U.S. Department of Energy through grant numbers DE-SC0020211, and DE-SC0024710. Electrical transport and XPS (at UMN) were supported by the Air Force Office of Scientific Research (AFOSR) through Grant No. FA9550-21-1-0025. Film growth was performed using instrumentation funded by AFOSR DURIP awards FA9550-18-1-0294 and FA9550-23-1-0085. S.N. was supported partially by the UMN MRSEC program under Award No. DMR-2011401. Parts of this work were carried out at the Characterization Facility, University of Minnesota, which receives partial support from the NSF through the MRSEC program under Award No. DMR-2011401. This work was supported by the National Research Foundation of Korea (NRF) grant funded by the Korea government (MSIT) (No., 2023R1A2C1005252, 2021R1A2C2011340, 2022M3H4A1A04074153, RS-2023-00220471, RS-2022-NR068223 (PAL 6A beamline), 2022R1A2C2007847 (I.H.C, S.w.L, and J.S.L), and RS-2023-00281671 (S.S. and S.P.). B.L. and J.M.L. acknowledge support from the AFOSR through grant No. FA9550-20-1-0066.

**Figures**

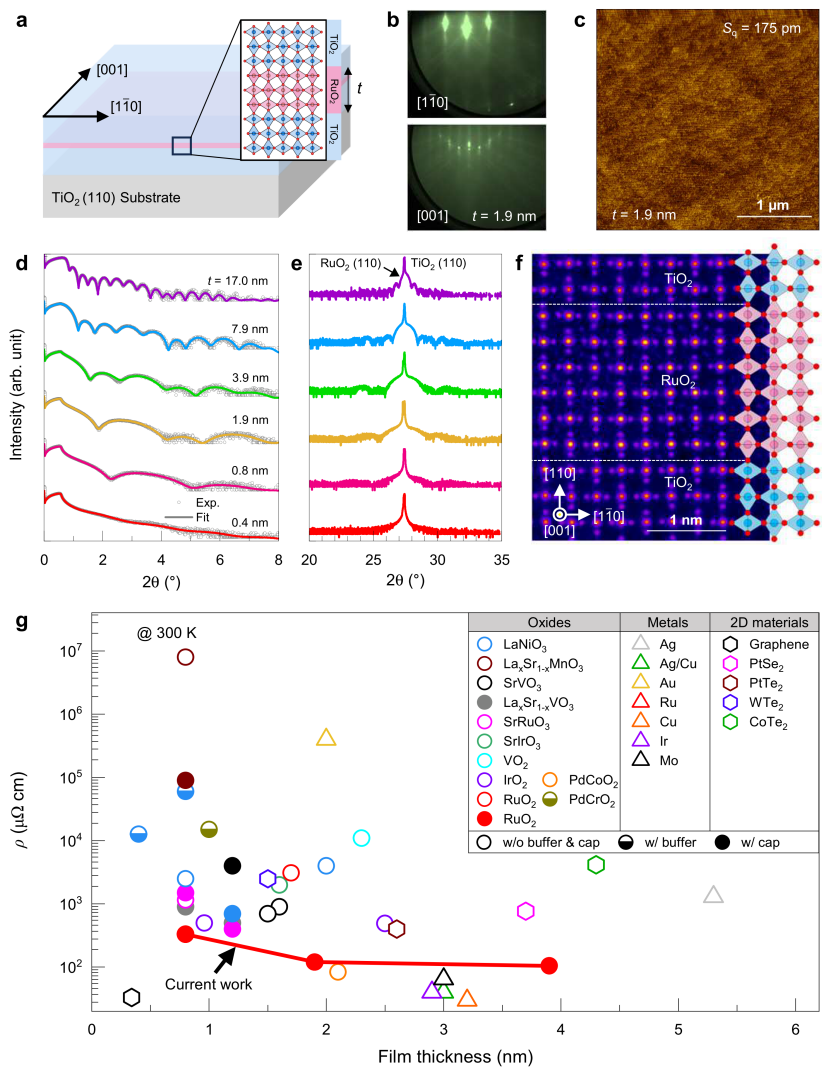

**Figure 1: Realization of metallicity in ultrathin RuO$_2$ (110) heterostructures using hMBE.** (a) Sample schematic comprising of 2 nm TiO$_2$/ $t$ nm RuO$_2$/ 2 nm TiO$_2$/ TiO$_2$ (110) heterostructure. (b) RHEED patterns after growth for $t$ = 1.9 nm film show the streaky RHEED patterns along [1$\bar{1}$0] (top) and [001] azimuthal directions (bottom). (c) AFM image for $t$ = 1.9 nm film showing atomically smooth surface topography with step-terrace structure. The root mean square roughness ($S_q$) is 175 pm. (d) XRR and (e) 2θ-θ XRD scans of RuO$_2$ heterostructures with different $t$. The scattered symbols and lines in XRR indicate experimental data and fitting results, respectively. (f) Cross-sectional electron ptychography image of 2 nm TiO$_2$/2.1 nm RuO$_2$/ 2 nm TiO$_2$ layers exhibiting atomically sharp interfaces. The right side displays an overlay with the corresponding rutile structures. (g) Comparison of room-temperature resistivity ($\rho$) of the samples as a function of $t$, including those of oxides (41 , 42-57), elemental metals (58-62), and 2D materials (63-67).

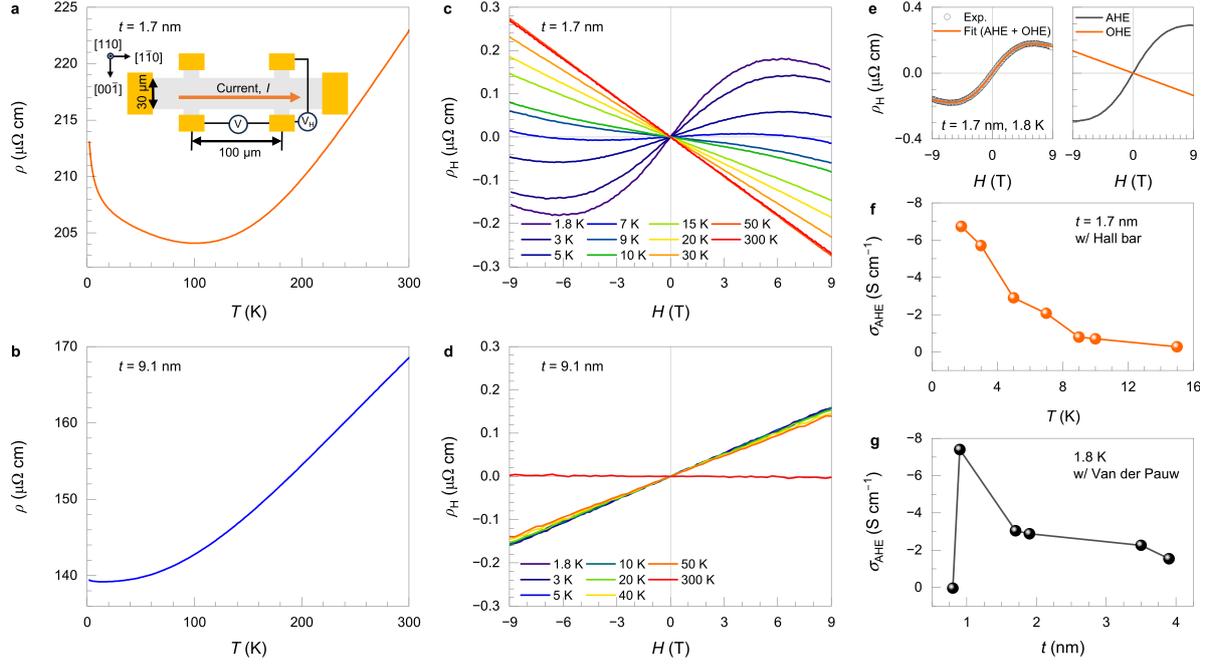

**Figure 2: Anomalous Hall effect in ultrathin-strained RuO₂ heterostructures.** Temperature-dependent $\rho$ of 2 nm TiO$_2$/ $t$ nm RuO$_2$/ 2 nm TiO$_2$/ TiO$_2$ (110) heterostructure for (a) $t$ = 1.7 nm and (b) 9.1 nm, measured by Hall bar devices. The inset in (a) shows the schematic of a Hall bar device. The $H$-field dependent Hall resistivity, $\rho_H$ for (c) $t$ = 1.7 and (d) 9.1 nm at selected temperatures between 1.8 and 300 K. (e) Fitting results of nonlinear $\rho_H(H)$ data (left panel) by combining AHE and OHE contributions (right panel). (f) Temperature-dependent $\sigma_{AHE}$ extracted from the fitting of data shown in (c). (g) $\sigma_{AHE}$ at 1.8 K as a function of $t$.

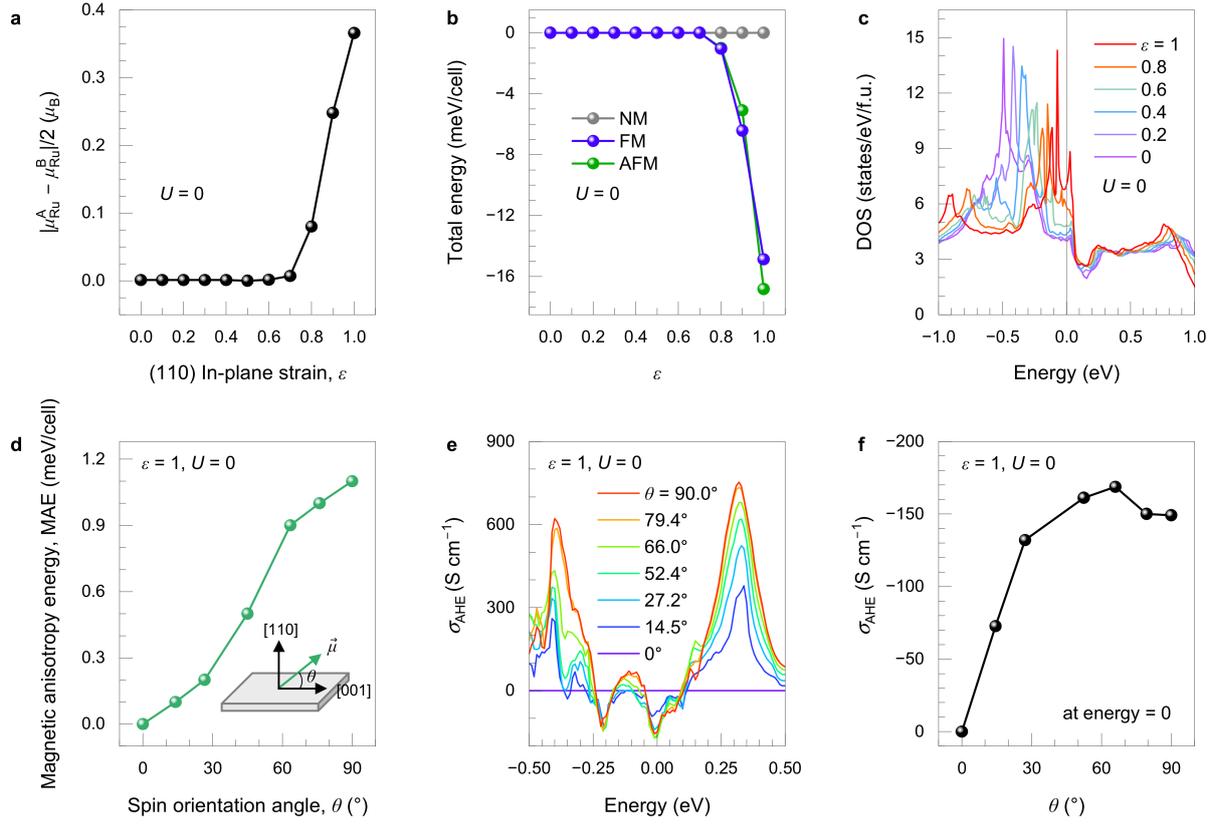

**Figure 3: DFT calculations of magnetic ground state and AHE in epitaxially-strained $RuO_2/TiO_2$ (110) heterostructures.** (a) The averaged local magnetic moment of Ru atoms, $|\mu_{Ru}^A - \mu_{Ru}^B|/2$ and (b) relative total energy as a function of the (110) in-plane epitaxial strain of $RuO_2$ at $U = 0$. (c) strain-dependent DOS near the Fermi level at $U = 0$ (d) Magnetic anisotropy energy (MAE) as a function of spin orientation angle ($\theta$) of a fully-strained $RuO_2$ film with $U = 0$. (e) Energy-dependent $\sigma_{AHE}$ of a fully-strained $RuO_2$ film with $U = 0$. (f) $\theta$-dependent $\sigma_{AHE}$ of a fully-strained $RuO_2$ at Fermi level (energy = 0 eV) and $U = 0$.